\documentclass[reprint,superscriptaddress,longbibliography]{revtex4-2}

\usepackage{amssymb}
\usepackage{amsmath}
\usepackage{hyperref}
\usepackage{graphicx}
\usepackage{url} 
\usepackage{enumitem}
\usepackage{textcomp}
\usepackage{gensymb}
\usepackage[export]{adjustbox}
\usepackage{array}
\usepackage{makecell}
\usepackage{listings}

\usepackage{ulem}

\begin{document}

\title[]{A Quantum Photonic Interface for Tin-Vacancy Centers in Diamond}

\author{Alison E. Rugar}
\thanks{These authors contributed equally to this work.}
\affiliation{E. L. Ginzton Laboratory, Stanford University, Stanford, CA 94305, USA}
\author{Shahriar Aghaeimeibodi}
\thanks{These authors contributed equally to this work.}
\affiliation{E. L. Ginzton Laboratory, Stanford University, Stanford, CA 94305, USA}
\author{Daniel Riedel}
\thanks{These authors contributed equally to this work.}
\affiliation{E. L. Ginzton Laboratory, Stanford University, Stanford, CA 94305, USA}
\author{Constantin Dory}
\affiliation{E. L. Ginzton Laboratory, Stanford University, Stanford, CA 94305, USA}
\author{Haiyu Lu}
\affiliation{Department of Physics, Stanford University, Stanford, California 94305, USA}
\affiliation{Geballe Laboratory for Advanced Materials, Stanford University, Stanford, California 94305, USA}
\affiliation{Stanford Institute for Materials and Energy Sciences, SLAC National Accelerator Laboratory, Menlo Park, California 94025, USA}
\author{Patrick J. McQuade}
\affiliation{Stanford Institute for Materials and Energy Sciences, SLAC National Accelerator Laboratory, Menlo Park, California 94025, USA}
\affiliation{Department of Materials Science and Engineering, Stanford University, Stanford, California 94305, USA}
\author{Zhi-Xun Shen}
\affiliation{Department of Physics, Stanford University, Stanford, California 94305, USA}
\affiliation{Geballe Laboratory for Advanced Materials, Stanford University, Stanford, California 94305, USA}
\affiliation{Stanford Institute for Materials and Energy Sciences, SLAC National Accelerator Laboratory, Menlo Park, California 94025, USA}
\affiliation{Department of Applied Physics, Stanford University, Stanford, California 94305, USA}
\author{Nicholas A. Melosh}
\affiliation{Stanford Institute for Materials and Energy Sciences, SLAC National Accelerator Laboratory, Menlo Park, California 94025, USA}
\affiliation{Department of Materials Science and Engineering, Stanford University, Stanford, California 94305, USA}
\author{Jelena Vu\v{c}kovi\'c}
\email{jela@stanford.edu}
\affiliation{E. L. Ginzton Laboratory, Stanford University, Stanford, CA 94305, USA}
\date{\today}

\begin{abstract}
The realization of quantum networks critically depends on establishing efficient, coherent light-matter interfaces. Optically active spins in diamond have emerged as promising quantum nodes based on their spin-selective optical transitions, long-lived spin ground states, and potential for integration with nanophotonics. Tin-vacancy (SnV$^{\,\textrm{-}}$) centers in diamond are of particular interest because they exhibit narrow-linewidth emission in nanostructures and possess long spin coherence times at temperatures above 1~K. However, a nanophotonic interface for SnV$^{\,\textrm{-}}$ centers has not yet been realized. Here, we report cavity enhancement of the emission of SnV$^{\,\textrm{-}}$ centers in diamond. We integrate SnV$^{\,\textrm{-}}$ centers into one-dimensional photonic crystal resonators and observe a 40-fold increase in emission intensity. The Purcell factor of the coupled system is 25, resulting in channeling of the majority of photons ($90\%$) into the cavity mode. Our results pave the way for the creation of efficient, scalable spin-photon interfaces based on SnV$^{\,\textrm{-}}$ centers in diamond.

\end{abstract}
\maketitle

\section{Introduction}
The basis of scalable quantum networks are nodes with optically accessible, long-lived quantum memories coupled to efficient photonic interfaces\,\cite{Kimble2008}.
A critical milestone toward the implementation of a quantum network is the development of an efficient, coherent light-matter interface\,\cite{Reiserer2015,Monroe2013}. Such an interface can be implemented by coupling optically active spin qubits to nanophotonic cavities. By strongly confining optical fields, nanocavities enhance the coherent emission of embedded qubits and channel the emitted photons into a single optical mode.
Diamond hosts a number of color centers that are excellent optically interfaced spin qubit candidates\,\cite{Awschalom2018,Atature2018,Janitz2020}. The most established of these color centers is the nitrogen-vacancy (NV$^{\,\textrm{-}}$) center, which has enabled seminal experiments such as heralded long-distance entanglement\,\cite{Hensen2015} and entanglement distillation\,\cite{Kalb2017}.
Unfortunately, fabrication-induced charge noise degrades the optical coherence of NV$^{\,\textrm{-}}$ centers and thus precludes the integration of NV$^{\,\textrm{-}}$ centers with nanophotonic cavities\,\cite{Faraon2012,Li2015}. The entanglement rates in these experiments are therefore limited by the small fraction of coherent emission into the NV$^{\,\textrm{-}}$ zero-phonon line (ZPL).

Group-IV color centers in diamond are significantly less sensitive to local charge fluctuations due to their inversion-symmetric structure\,\cite{Bradac2019}. As a consequence, these color centers exhibit nearly lifetime-limited linewidths in nanostructures despite the use of invasive plasma etching techniques\,\cite{Wan2020,Trusheim2020,Rugar2020b}.
While silicon-vacancy (SiV$^{\,\textrm{-}}$) centers coupled to nanophotonic resonators have enabled the demonstration of high-fidelity single-shot readout and nuclear spin memory-enhanced quantum communication\,\cite{Bhaskar2020}, SiV$^{\,\textrm{-}}$ centers suffer from low quantum efficiency, and their excellent spin properties are accessible only under high strain\,\cite{Sohn2018} or at millikelvin temperatures\,\cite{Sukachev2017}. Fortunately, other group-IV color centers in diamond are expected to share many of the favorable properties of SiV$^{\,\textrm{-}}$ centers but at higher temperatures\,\cite{Thiering2018}. Tin-vacancy (SnV$^{\,\textrm{-}}$) centers\,\cite{Iwasaki2017,Trusheim2020,Rugar2019,Gorlitz2020} stand out because of their high quantum efficiency\,\cite{Iwasaki2017} and long spin coherence times at cryogenic temperatures above 1~K\,\cite{Trusheim2020}. Despite their promise as optically active spin qubit candidates, the critical step of incorporating SnV$^{\,\textrm{-}}$ centers into cavities has yet to be realized.

In this letter, we report the coupling of the ZPL emission of SnV$^{\,\textrm{-}}$ centers to one-dimensional photonic crystal cavities. We fabricate a large array of devices on a single chip using a fabrication technique based on the quasi-isotropic diamond undercut method\,\cite{Khanaliloo2015,Mouradian2017,Wan2018,Mitchell2019,Dory2019}. 
By tuning the cavity into resonance with the color center, we demonstrate a 40-fold increase in SnV$^{\,\textrm{-}}$ center emission intensity. Furthermore, we observe a 10-fold reduction in the SnV$^{\,\textrm{-}}$ center excited-state lifetime, corresponding to a Purcell factor of $25$.
Because of this enhancement, the majority of photons are emitted into the cavity mode via the ZPL ($\beta=90\%$). With their excellent optical properties\,\cite{Iwasaki2017} and competitive spin coherence times accessible without a dilution refrigerator\,\cite{Trusheim2020}, SnV$^{\,\textrm{-}}$ centers integrated with nanocavities constitute a promising platform for quantum networks.

\begin{figure*}[htbp]
\includegraphics[width=0.8\textwidth,]{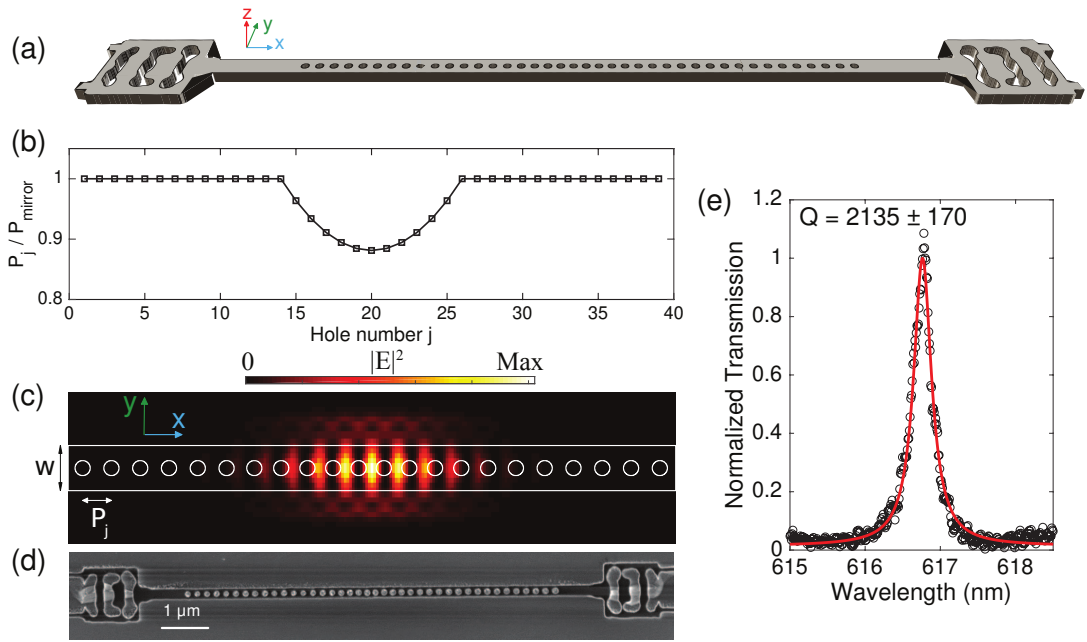}
\caption{Design and simulation of the nanobeam photonic crystal cavity. \textbf{(a)} Schematic of the proposed one-dimensional photonic crystal cavity. Two inverse-designed vertical couplers on either end of the beam enable the coupling of light into and out of the device. \textbf{(b)} Spacing between the holes for different hole numbers. A quadratic taper over the center 12 holes creates the cavity. \textbf{(c)} FDTD simulation of electric field magnitude squared $|E|^2$ showing confinement of light in the center of the cavity for the circular-hole design. \textbf{(d)} Scanning electron microscope image of a fabricated device. \textbf{(e)} Spectrum of a cavity mode of a representative device obtained from a transmission measurement through the device. The quality factor of this device is $Q=2135 \pm 170 $.}
\label{device_fig}
\end{figure*}

\section{Results}
Our cavities are based on one-dimensional photonic crystals where a periodic array of circular holes etched into a suspended diamond waveguide creates a photonic bandgap\,\cite{Gong2010}. 
As illustrated in Fig.~\ref{device_fig}(a), we implement inverse-designed vertical couplers on either end of the cavity to facilitate the in- and out-coupling of the light\,\cite{Dory2019}. 
Fig.~\ref{device_fig}(b) shows the designed spacing between the holes for a device with 20 holes on each side. 
We quadratically taper the spacing between the 12 central holes from $P_\textrm{mirror}$ in the mirror section to a reduced period $P_\textrm{center}= 0.88~ P_\textrm{mirror}$ in the center of the cavity to confine the light.
We select the waveguide width ${w=300}$~nm, thickness $h=200$~nm, period $P_\textrm{mirror}=190$~nm, and hole radius $r= 57$~nm in order to have a resonance around 620~nm of high quality factor ($Q$) and small mode volume ($V_\textrm{mode}$). 
We simulate the performance of our design using three-dimensional Finite-Difference Time-Domain method (FDTD; Lumerical).
Fig.~\ref{device_fig}(c) illustrates the simulated $|E|^2$ for the above parameters with a simulated $Q= 2\times 10^5$ and $V_\textrm{mode}= 0.56~(\lambda/n)^3$.

Next, we fabricate our photonic crystal cavities. We first generate SnV$^{\,\textrm{-}}$ centers at a depth of $90~\textrm{nm}$ with our recently developed shallow ion implantation and growth method\,\cite{Rugar2020}. We then perform a fabrication routine based on quasi-isotropic etching\,\cite{Khanaliloo2015,Mouradian2017,Wan2018,Mitchell2019,Dory2019} to create a large matrix of devices. In order to account for shifts of the cavity resonance wavelength caused by fabrication imperfections, we vary $P_\textrm{mirror}$ by $\pm20\%$. We also vary the number of mirror periods. Fig.~\ref{device_fig}(d) displays a scanning electron microscope (SEM) image of a representative fabricated cavity. Further details on color center generation and device fabrication are provided in the Methods section.

\begin{figure*}[htbp]
\includegraphics[width=0.8\textwidth,]{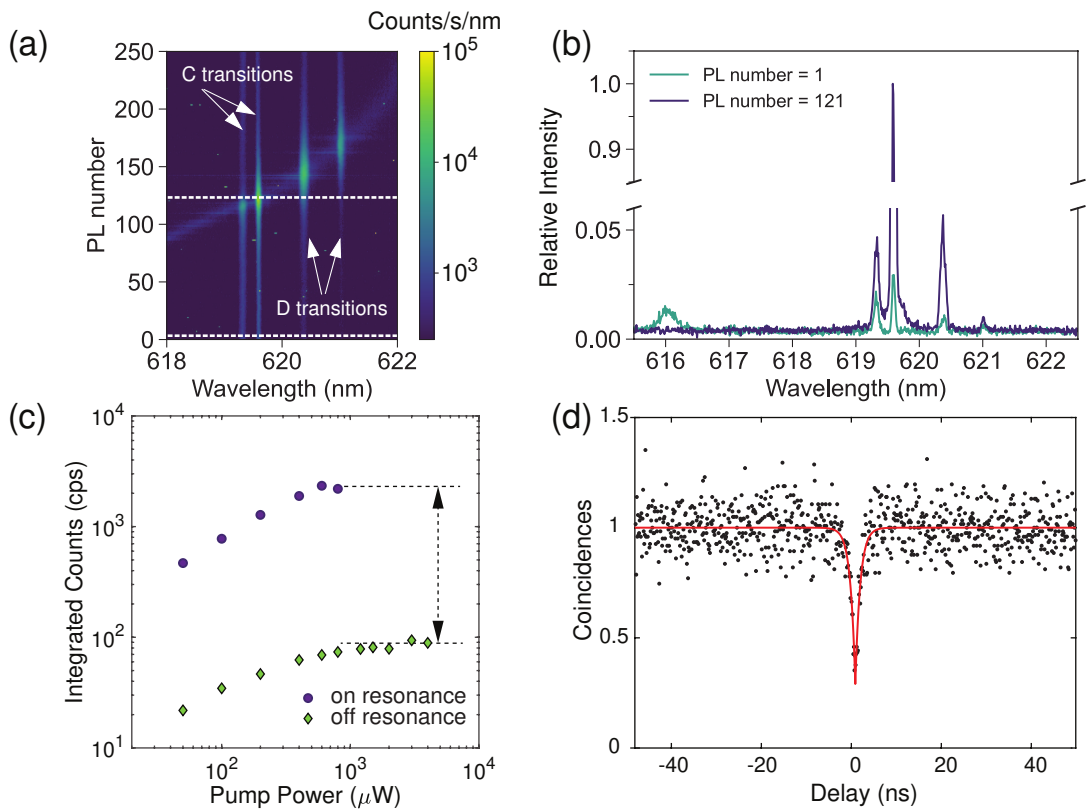}
\caption{Intensity enhancement of SnV$^{\,\textrm{-}}$ centers. 
\textbf{(a)} Consecutive PL spectra acquired while gas tuning the cavity through resonance with the ZPLs of the two SnV$^{\,\textrm{-}}$ centers. Two horizontal white dashed lines indicate locations of line cuts examined in panel (b). \textbf{(b)} Two line cuts acquired when the cavity mode is 3.6~nm blue-detuned from the C transition (teal) and when the cavity mode is resonant with the C transition of interest at 619.6~nm (purple). \textbf{(c)} Saturation of the SnV$^{\,\textrm{-}}$ center when cavity mode is 2.72 nm blue-detuned from the resonance (green diamonds) and when the cavity mode is resonant with the C transition at 619.6~nm (purple circles). In the resonant case the maximum achieved intensity of the C transition of the SnV$^{\,\textrm{-}}$ center is about 30 times greater than when the cavity is off resonance. \textbf{(d)} Second-order autocorrelation measurement of SnV$^{\,\textrm{-}}$ center when the cavity mode is resonant with the C transition. A fit (red) to the data (black) reveals a $g^{(2)}(0)=0.29 \pm 0.08$.}
\label{intensity_fig}
\end{figure*}

We characterize our devices at cryogenic temperatures (Montana Instruments, $\sim 5$~K) in a home-built confocal microscope setup. We look for devices that have a cavity mode that is slightly blue-shifted  with respect to the optical transitions of the SnV$^{\,\textrm{-}}$ center so that we can eventually use gas tuning to tune the cavity mode into resonance with the SnV$^{\,\textrm{-}}$ center ZPLs (see Methods). We also check the devices for SnV$^{\,\textrm{-}}$ center photoluminescence (PL) signal.

 To determine the quality factor and resonance wavelength of our cavities, we couple a supercontinuum laser through the vertical couplers and perform a broadband transmission measurement. Fig.~\ref{device_fig}(e) shows the measured transmission spectrum of a suitable device (${P_{\textrm{mirror}} = 228}$~nm  and 14 holes), featuring a quality factor of $2135 \pm 170$ at $616.76~\textrm{nm}$.

Next, we perform PL spectroscopy (see Methods) on the device to verify the presence of SnV$^{\,\textrm{-}}$ centers in the device. 
Fig.~\ref{intensity_fig}(a) displays a heatmap of consecutive PL spectra. The four vertical lines apparent in Fig.~\ref{intensity_fig}(a) are the ZPLs of two SnV$^{\,\textrm{-}}$ centers present in the cavity. At 5~K, individual SnV$^{\,\textrm{-}}$ centers have two prominent ZPLs centered about 620~nm, known as the C and D transitions\,\cite{Iwasaki2017}, where C transitions are higher in energy than D transitions. For the remainder of this paper, we focus on the C transition located at 619.6~nm, which has the highest count rate.

We characterize the intensity of the C transition at different detunings between the cavity mode and the C transition.
In Fig.~\ref{intensity_fig}(a), the cavity mode starts blue-detuned from the SnV$^{\,\textrm{-}}$ center ZPLs. The cavity mode is then tuned through the ZPLs of the SnV$^{\,\textrm{-}}$ centers by argon gas condensation. We examine in Fig.~\ref{intensity_fig}(b) two spectra from Fig.~\ref{intensity_fig}(a) corresponding to the white dashed lines: one spectrum in which the cavity is 3.6~nm blue-detuned from the C transition of interest (PL number~$= 1$) and another in which the cavity mode is resonant with the C transition at 619.6~nm (PL number~$= 121$). 
Between the off- and on-resonance cases shown in Fig.~\ref{intensity_fig}(b), the intensity of the C transition increases by a factor of $40\pm4$, indicating that this SnV$^{\,\textrm{-}}$-cavity system has a large Purcell factor. The intensity enhancements of the other C transition at 619.3~nm and the D transitions at 620.4~nm and 621.0~nm are respectively $11\pm2$, $56\pm11$, and $42\pm17$.

To further examine the intensity enhancement, we characterize the emission intensity of the C transition as a function of excitation power when the cavity mode is off- and on-resonance with the C transition. The resulting saturation data are presented in Fig.~\ref{intensity_fig}(c). We observe up to 30 times greater intensity for the resonant case than for the off-resonance case. At high excitation powers, the deposited Ar starts to evaporate and the cavity becomes detuned from the C transition. This effect limits the range of excitation powers that we can use without removing Ar to $\sim1~\textrm{mW}$. This large intensity contrast combined with the 40-fold intensity enhancement observed in Fig.~\ref{intensity_fig}(b) indicates that the SnV$^{\,\textrm{-}}$-cavity system under study has a large Purcell factor, which we will later quantify.

\begin{figure*}[htbp]
\includegraphics[width=0.8\textwidth,]{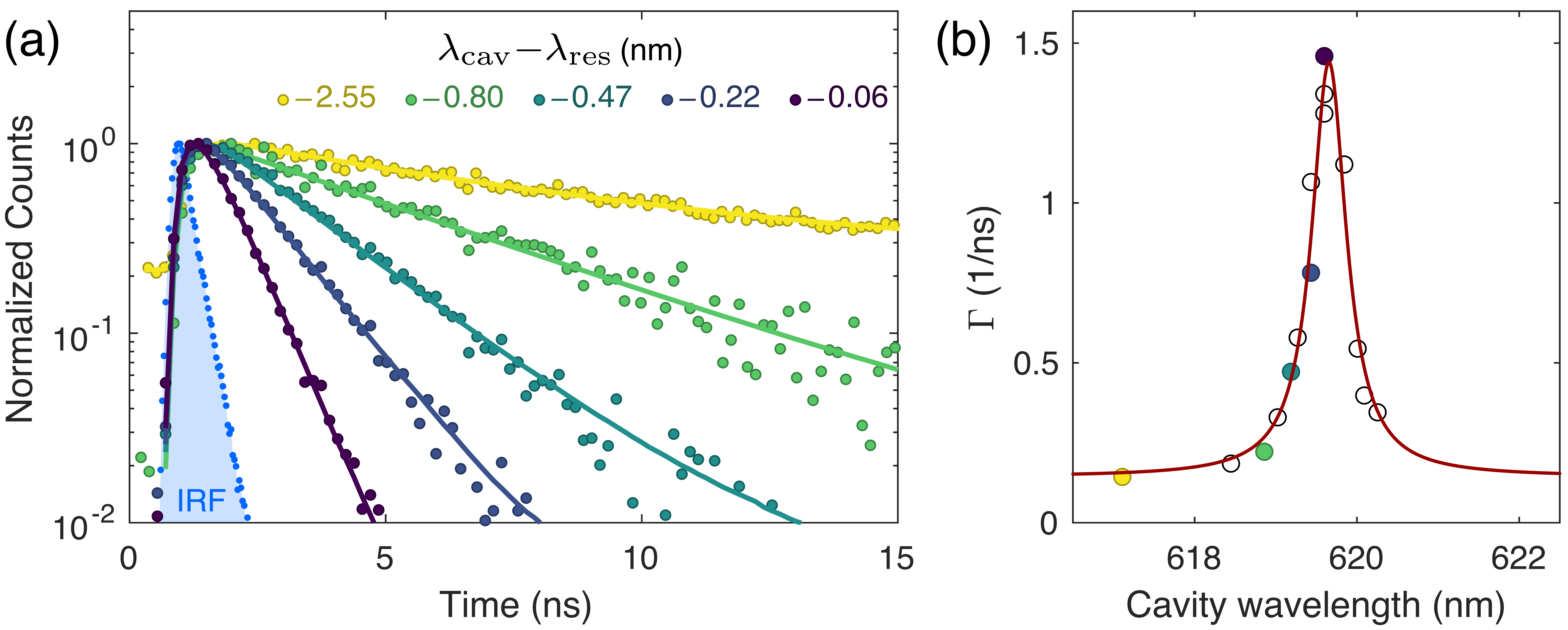}
\caption{Time-resolved PL measurements. \textbf{(a)} Representative lifetime measurements at different cavity wavelengths. To fit the data we convolve a single-exponential decay with the instrument response function (IRF) of our detectors (blue). \textbf{(b)} PL recombination rates of the studied SnV$^{\,\textrm{-}}$ center plotted as a function of cavity resonance wavelength ($\Gamma(\lambda_\textrm{cav})$). The data (circles) are fit with a Lorentzian model (red curve) to quantify the enhancement provided by the cavity. Filled circles correspond to lifetime data of the same color shown in (a). }
\label{lifetime_fig}
\end{figure*}

We confirm that the enhanced SnV$^{\,\textrm{-}}$ center is a single-photon source by performing a Hanbury-Brown-Twiss experiment with the cavity mode resonant with the C transition. The resulting second-order autocorrelation data, shown in Fig.~\ref{intensity_fig}(d), displays a distinct antibunching dip. A fit to the data with a function of the form $g^{(2)}(\tau)=1-\left(1-g^{(2)}(0)\right)e^{-|\tau|/\tau_0}$, where $\tau$ is the delay between detection events, reveals a  $\tau_0 = 1.09 \pm 0.2$~ns and $g^{(2)}(0)=0.29\pm0.08$, indicating that the enhanced C transition is emission from a single quantum emitter.

To quantify the Purcell enhancement of the system, we measure the excited-state lifetime of the SnV$^{\,\textrm{-}}$ center for different cavity resonance wavelengths ($\lambda_\textrm{cav}$) through time-resolved PL measurements (see Methods). To tune the cavity resonance wavelength, we inject Ar gas and determine $\lambda_{\textrm{cav}}$ using a reflectivity measurement prior to each lifetime measurement.
 Fig.~\ref{lifetime_fig}(a) displays representative lifetime measurements at different cavity wavelengths integrated for 100~s. To remove background contributions we subtract an off-resonant measurement ($\lambda_\textrm{cav}=616.1~\textrm{nm}$) acquired under the same experimental conditions. We fit the data to a single exponential decay convolved with the instrument response function (IRF) of our detectors.
 We then extract the excited-state decay rate $\Gamma(\lambda_\textrm{cav})$ from the respective lifetime fits.
A comprehensive study of the effect of the cavity on the PL decay rates of the SnV$^{\,\textrm{-}}$ center is shown in Fig.~\ref{lifetime_fig}(b). 
 To get a better estimate for the non-resonant radiative decay rate $\Gamma_{\textrm{off}}$, we perform an additional measurement at $\lambda_\textrm{cav}=617.1~\textrm{nm}$ with an increased integration time (1000~s) and use a single-mode fiber for detecting the PL. We find $\tau_{617.1~\textrm{nm}}=(6.980\pm0.078)~\textrm{ns}$, which is comparable to previously reported lifetimes for SnV$^{\,\textrm{-}}$ centers close to the diamond bulk surface\,\cite{Gorlitz2020}.
When the cavity mode is resonant with the C transition, the lifetime is strongly reduced to $\tau_{619.6~\textrm{nm}}=(0.685\pm0.014)~\textrm{ns}$.

We determine the relevant system parameters by fitting the data in Fig.~\ref{lifetime_fig}(b) to a Lorentzian model\,\cite{Riedel2017} $$\Gamma(\lambda_\textrm{cav})=\Gamma_{\textrm{off}}+\Gamma_{\textrm{cav}}\delta \lambda_\textrm{cav}^2/
\left(
{4(\lambda_\textrm{cav}-\lambda_\textrm{res})^2 +\delta \lambda_\textrm{cav}^2}
\right).$$ 
Here, the off-resonance decay rate is fixed to ${\Gamma_{\textrm{off}}=\Gamma_{617.1~\textrm{nm}}=0.143~\textrm{ns}^{-1}}$. We find a cavity-induced decay rate of ${\Gamma_{\textrm{cav}}=(1.30\pm0.16)~\textrm{ns}^{-1}}$ corresponding to an on-resonance decay rate of ${\Gamma_{\textrm{on}}=\Gamma_{\textrm{off}}+\Gamma_{\textrm{cav}}=(1.44\pm0.16)~\textrm{ns}^{-1}}$. The width and center wavelength of the Lorentzian fit are ${\delta\lambda_\textrm{cav}=(0.260\pm0.040)~\textrm{nm}}$ and ${\lambda_\textrm{res}=(619.656\pm0.032)~\textrm{nm}}$, respectively.

On resonance, the overall lifetime reduction caused by the interaction of the SnV$^{\,\textrm{-}}$ center with the cavity is given by $\tau_{\textrm{off}}/\tau_{\textrm{on}}=\Gamma_{\textrm{on}}/\Gamma_{\textrm{off}}=10.1\pm1.2$.
From the cavity-induced decay rate $\Gamma_{\textrm{cav}}=F_\textrm{P}^\textrm{exp}\xi_\textrm{c}\Gamma_{\textrm{off}}$ we determine the experimental Purcell factor $F_\textrm{P}^\textrm{exp}=24.8\pm3.0$. 
Here, $\xi_\textrm{c}=0.36$ is a factor that corrects for the non-unity probability of the SnV$^{\,\textrm{-}}$ center relaxing radiatively via the C transition. $\xi_\textrm{c}$ is a product of quantum efficiency ($80\%$\,\cite{Iwasaki2017}), Debye-Waller factor ($57\%$\,\cite{Gorlitz2020}), and branching ratio into the C transition ($80\%$, extracted from an off-resonance PL spectrum of this emitter).
The figure of merit of our system, $\beta$, the resulting probability of an excited-state decaying through emission into the cavity mode via the C transition, is approaching unity: $\beta=\Gamma_{\textrm{cav}}/\Gamma_{\textrm{on}}=(90.1\pm1.1)\%$.

To benchmark our system performance we calculate the theoretical Purcell enhancement for an emitter positioned at the maximum cavity field.
The Lorentzian fit to $\Gamma(\lambda_\textrm{cav})$ yields a cavity $Q$ factor of $2384\pm366$ (Fig.~\ref{lifetime_fig}(b)) which is consistent with transmission measurement in Fig.~\ref{device_fig}(e). The theoretical Purcell factor of our device is thus $F_\textrm{P}^\textrm{theo}=3/(4\pi^2)~Q/V_\textrm{mode}~(\lambda/n)^3=302\pm46.$
Accounting for the angular mismatch between cavity polarization of $\langle110\rangle$ and SnV$^{\,\textrm{-}}$ center symmetry axis along $\langle111\rangle$ ($\cos^2{\phi}=2/3$), we infer that the studied SnV$^{\,\textrm{-}}$ experiences only $(12.3\pm2.4)\%$ of the maximum cavity field $|E_\textrm{max}|^2$. This discrepancy hints at a displacement of the studied emitter from the field maximum.

\section{Conclusion}
We have demonstrated the Purcell enhancement of a SnV$^{\,\textrm{-}}$ center in a diamond photonic crystal cavity. With our SnV$^{\,\textrm{-}}$-cavity system, we can achieve a 40-fold increase in emission intensity into the C transition. The system displays a lifetime reduction of 10 and a Purcell factor of 25. As a result, $90\%$ of the PL emission is channeled through the ZPL into the cavity mode, enabling a ZPL photon creation rate in excess of 1~GHz. The Purcell factor could be further increased by improving the $Q$/$V_\textrm{mode}$ ratio of the cavities\,\cite{Mouradian2017,Mitchell2019}. Additionally, deterministic positioning of the emitter can improve the yield of devices with large Purcell enhancement. Sub-10-nm placement accuracy can be achieved by combining our shallow-ion implantation and growth technique\,\cite{Rugar2020} with implantation masks\,\cite{Toyli2010,Bayn2015,Scarabelli2016}. 
For their use in extended quantum networks, SnV$^{\,\textrm{-}}$ center ZPL photons need to be efficiently coupled into an optical fiber network\,\cite{Wehner2018}.
To that end, the main loss channel of the coupled system needs to be transmission into the waveguide mode. Waveguide-to-fiber coupling efficiencies of \textgreater$90\%$ can be achieved using fiber tapers\,\cite{Burek2017}. As an alternative, optimized inverse-designed vertical couplers have the advantage of featuring a small footprint while potentially exhibiting similar coupling efficiencies \textgreater$85\% $\,\cite{Dory2019}.
Combining our SnV$^{\,\textrm{-}}$-cavity systems with on-chip photonic architectures via hybrid integration techniques would enable large-scale quantum information processing systems \,\cite{Elshaari2020,Kim2020,Wan2020}. Our work paves the way toward establishing a coherent and efficient spin-photon interface based on diamond color centers without the need for dilution refrigerators.

\section{Methods}
\subsection{Device fabrication}
We fabricate our nanophotonic resonators from an electronic-grade single-crystalline diamond plate (Element Six). The chip is first cleaned in a boiling tri-acid solution (1:1:1 sulfuric/nitric/perchloric acids). We then remove the top 500~nm of the chip with an oxygen (O$_2$) plasma etch. By employing our recently developed shallow ion implantation and growth (SIIG) method\,\cite{Rugar2020} we create a $\delta$-doped layer of high-quality SnV$^{\,\textrm{-}}$ centers. Here, $^{120}$Sn$^+$ ions are implanted shallowly using low implantation energies (1~keV) with a dose of $5\times10^{11}$~cm$^{-2}$. Ion implantation was performed by CuttingEdge Ions.
A thin film (90~nm) of high-quality diamond material is subsequently grown by microwave-plasma chemical vapor deposition (Seki Diamond Systems SDS 5010; 300~sccm H$_2$, 0.5~sccm CH$_4$, stage temperature of 650\degree\,C, microwave power of 1100~W, and pressure of 23~Torr).

We fabricate our photonic devices via the quasi-isotropic etching technique\,\cite{Khanaliloo2015,Mouradian2017,Wan2018,Mitchell2019,Dory2019}.
First, 200~nm of Si$_x$N$_y$ are grown via plasma-enhanced chemical vapor deposition. The structures are then patterned in hydrogen silsesquioxane FOx-16 via electron-beam lithography. The Si$_x$N$_y$ is then etched with SF$_6$, CH$_4$, and N$_2$ reactive ion etch (RIE). We use the patterned Si$_x$N$_y$ layer as an etch mask for the diamond substrate. The diamond is etched with an anisotropic O$_2$ RIE. We then grow 30~nm of Al$_2$O$_3$ via atomic layer deposition. The horizontal planes of the Al$_2$O$_3$ layer are removed with a Cl$_2$, BCl$_2$, and N$_2$ RIE so that only the sidewalls of the diamond structures are covered by Al$_2$O$_3$. Using a second anisotropic O$_2$ RIE we expose bare diamond sidewalls. The quasi-isotropic etch\,\cite{Khanaliloo2015,Mouradian2017,Wan2018,Mitchell2019,Dory2019} step is now performed to undercut the structures. This O$_2$ plasma etch step is performed at high temperature (300\degree\,C) with zero forward bias and high inductively coupled plasma power\,\cite{Khanaliloo2015,Mouradian2017,Wan2018,Mitchell2019,Dory2019}. This etch progresses preferentially along the $\{110\}$ planes\,\cite{Mouradian2017}. Once the nanobeam waveguides have been released and etched to the desired thickness, as validated by high-voltage SEM, the sample is soaked in hydrofluoric acid to remove the Si$_x$N$_y$ and Al$_2$O$_3$ etch masks.

\subsection{Measurements}
All measurements were conducted in a home-built confocal microscope setup, with the sample cooled to $\sim5$~K in a Montana Instruments Cryostation.

To tune the cavity mode, we employ gas tuning. Argon gas is injected into the cryostat and freezes onto the devices. Condensed Ar increases the effective refractive index of the device and thus red-shifts the cavity resonance wavelength. Gas deposition can be reversed by heating, either locally with a laser or chip-wide with a heated stage. 

For all PL measurements presented in Fig.~\ref{intensity_fig}, a continuous-wave, 532-nm laser filtered by a $532\pm3$-nm bandbass filter is used to excite the SnV$^{\,\textrm{-}}$ centers. For the PL data presented in Fig.~\ref{intensity_fig}(a), the excitation power was set to $500~\mu\textrm{W}$. Before being coupled into a single-mode fiber, the collected light is filtered by a 568-nm long-pass filter and a 532-nm notch filter. The excitation and collection spots are aligned to the same spot, on top of the center of the cavity.

For the $g^{(2)}$ measurement, the excitation power was set to $450~\mu\textrm{W}$. The light collected by the single-mode fiber is filtered by a monochromator (Princeton Instruments Acton SP2750). The light is then sent via multi-mode fiber to a Hanbury-Brown-Twiss setup comprising a fiber-beamsplitter and two single-photon counting modules (SPCM; PerkinElmer SPCM-AQR-14-FC). A time-correlated single photon counting module (TCSPC; PicoHarp 300) was used to construct the histogram shown in Fig.~\ref{intensity_fig}(d).

To measure the lifetimes of the SnV$^{\,\textrm{-}}$ center in the cavity presented in Fig.~\ref{lifetime_fig}, we used a supercontinuum laser (Fianium SC400) filtered with a 450-nm-long-pass filter and a 550-nm short-pass filter to excite the SnV$^{\,\textrm{-}}$ center. The emitted light was then filtered by a 568-nm long-pass filter before being coupled into a fiber. The collected light was further filtered by a monochromator (Princeton Instruments Acton SP2750) before being coupled into a multi-mode fiber and sent to a SPCM. The lifetime measurements were recorded using the TCSPC.

\section{Acknowledgments}
This work is financially supported by Army Research Office (ARO) (award no. W911NF-13-1-0309); National Science Foundation (NSF) RAISE TAQS (award no. 1838976); Air Force Office of Scientific Research (AFOSR) DURIP (award no. FA9550-16-1-0223). Stanford Institute for Materials and Energy Sciences (SIMES) research is supported by the Division of Materials Science and Engineering, Office of Basic Energy Sciences, DOE, and SLAC LDRD. A.E.R. acknowledges support from the National Defense Science and Engineering Graduate (NDSEG) Fellowship Program, sponsored by the Air Force Research Laboratory (AFRL), the Office of Naval Research (ONR) and the Army Research Office (ARO). S.A. acknowledges support from Bloch postdoctoral fellowship in quantum science and engineering from Stanford Q-FARM. D.R. acknowledges support from the Swiss National Science Foundation (Project P400P2\_194424). C.D. acknowledges support from the Andreas Bechtolsheim Stanford Graduate Fellowship and the Microsoft Research PhD Fellowship. Part of this work was performed at the Stanford Nanofabrication Facility (SNF) and the Stanford Nano Shared Facilities (SNSF), supported by the National Science Foundation under award ECCS-2026822.


%

\end{document}